\documentclass[11pt, a4paper]{article}
\usepackage{amsmath}
\newtheorem{theorem}{Theorem}

\begin{document}

\title{Classification of integrable Hamiltonian hydrodynamic chains associated with Kupershmidt's brackets}
\author{E.V. Ferapontov, K.R. Khusnutdinova, D.G. Marshall and M.V. Pavlov
   \footnote{On the leave from: Lebedev Physical Institute, Academy of
   Sciences of Russia, Moscow,  Russia.}}
    \date{}
    \maketitle
    \vspace{-7mm}
\begin{center}
Department of Mathematical Sciences \\ Loughborough University \\
Loughborough, Leicestershire LE11 3TU \\ United Kingdom \\[2ex]
e-mails: \\[1ex] \texttt{E.V.Ferapontov@lboro.ac.uk}\\
\texttt{K.R.Khusnutdinova@lboro.ac.uk}\\
\texttt{D.G.Marshall@lboro.ac.uk}\\
\texttt{M.V.Pavlov@lboro.ac.uk}

\end{center}

\bigskip

\begin{abstract}
We characterize a class of integrable Hamiltonian hydrodynamic chains, based on the necessary condition for the integrability provided by the vanishing of the Haantjes tensor. We prove that the vanishing of the first few components of the Haantjes tensor is already sufficiently restrictive, and allows a complete description of the corresponding Hamiltonian densities. In each of the cases  we were able to explicitly construct a generating function for conservation laws, thus establishing the integrability.

\bigskip

\noindent MSC: 35L40, 35L65, 37K10.

\bigskip

Keywords: Hamiltonian hydrodynamic chains, Integrability, Haantjes tensor, Reductions,
Conservation laws.

\end{abstract}

\section{Introduction}

We consider Hamiltonian systems of the form
\begin{eqnarray}
 {\bf u}_t =\left (B\frac{d}{dx}+\frac{d}{dx}B^t\right )\frac{\partial h}{\partial {\bf u}}
 \label{Ham}
 \end{eqnarray}
 where ${\bf u}=(u^1,u^2,u^3, ...)^t$ is an infinite-component column vector of the dependent variables, and  the matrix $B$ of the Hamiltonian operator $B\frac{d}{dx}+\frac{d}{dx}B^t$ is defined as $B^{ij}=(\alpha(i-1)+\beta)u^{i+j-1}$. Explicitly, one has
\begin{eqnarray*}
B=\left( \begin{array}{cccc}
\beta u^1 & \beta u^2 & \beta u^3 & ....\\
(\alpha+\beta) u^2 &(\alpha+\beta) u^3 & (\alpha+\beta)  u^4 &....\\
(2\alpha+\beta) u^3 &(2\alpha+\beta) u^4 &(2\alpha+\beta) u^5 & ....\\
.... & .... & .... & ....\\
\end{array} \right);
\end{eqnarray*}
here $\alpha$ and $\beta$ are arbitrary constants. Hamiltonian operators of this type first appeared in \cite{Kup3}, and belong to the general class introduced in  \cite{Dorfman}. The Hamiltonian density $h$ is assumed to be a function of the first two independent variables $u^1,u^2$ only: $h=h(u^1,u^2)$. In this paper we address the problem of the classification of all densities $h(u^1, u^2)$ such that the corresponding Hamiltonian system is {\it integrable} (partial  results were reported earlier  in \cite{Kup3}). The integrability of  Hamiltonian chains of the type (\ref{Ham}) can be defined by either of the two  properties:

\medskip

\noindent (1) The existence of infinitely many additional 
conservation laws which Poisson commute with the Hamiltonian $\int h(u^1, u^2) \ dx$, see \cite{Kup3}.

\medskip

\noindent (2) The existence of infinitely many  hydrodynamic reductions, see \cite{GibTsa96, GibTsa99, FerMar}.

\medskip

To derive the integrability conditions we utilize  the second approach, based on the calculation of the so-called Haantjes tensor \cite{Haantjes}. Let us first represent the system (\ref{Ham}) in a hydrodynamic form,
$$
 {\bf u}_t = V({\bf u})\ {\bf u}_x,
$$
 where $V=v^i_j$ is an $\infty \times \infty$ matrix. This matrix has the following  properties:
 
 \noindent (a) each row of $V$ contains {\it finitely many} nonzero elements;

\noindent (b) each matrix element of $V$ depends on {\it finitely many} variables $u^i$.

\noindent Infinite systems of this type are known as `hydrodynamic chains'. 
The properties (a) and (b) ensure that  both the Nijenhuis tensor,
\begin{equation}
N^i_{jk}=v^p_j\partial_{u^p}v^i_k-v^p_k\partial_{u^p}v^i_j-v^i_p(\partial_{u^j}v^p_k-\partial_{u^k}v^p_j),
\label{N}
\end{equation}
(the standard summation convention over  repeated indices is adopted), and  the Haantjes tensor,
\begin{equation}
H^i_{jk}=N^i_{pr}v^p_jv^r_k-N^p_{jr}v^i_pv^r_k-N^p_{rk}v^i_pv^r_j+N^p_{jk}v^i_rv^r_p,
\label{H}
\end{equation}
are well-defined objects, so that the calculation of each particular component $H^i_{jk}$ requires finitely many summations only. Moreover, for a fixed upper index $i$, one has {\it finitely many} nonzero components $H^i_{jk}$, see \cite{FerMar}. 
 According to the results of Tsarev \cite{Tsarev},  the vanishing of the Haantjes tensor is  necessary and sufficient for the  integrability of finite-component Hamiltonian systems of hydrodynamic type by the generalized hodograph method. Thus, we formulate our main

\medskip

\noindent {\bf Conjecture.} {\it The vanishing of the Haantjes tensor is a necessary and sufficient condition for the integrability of Hamiltonian hydrodynamic chains. In particular, it implies the existence of infinitely many Poisson commuting conservation laws, and infinitely many hydrodynamic reductions.}

\medskip

The necessity part of this conjecture  follows from the general result of  \cite{FerMar}, according to which  the vanishing of the Haantjes tensor is a necessary condition for  the integrability of hydrodynamic chains (not necessarily Hamiltonian). The sufficiency is a far more delicate property and requires, as a sub-problem, the classification of infinite-component Hamiltonian structures. The conjecture is supported by all examples of integrable Hamiltonian chains known to us. 
Since components of the Haantjes tensor can be calculated using computer algebra, this approach provides an effective classification criterion.  The main goal of this paper is to demonstrate  that the conjecture is indeed true for Hamiltonian chains of the type (\ref{Ham}).

Upon setting the first components $H^1_{jk}$  equal to zero we obtain the expressions for all  third order partial derivatives of the Hamiltonian density $h(u^1,u^2)$ in terms of  lower order derivatives, see equations (\ref{h}) in Sect 2. A complete list of integrable Hamiltonian densities is obtained in Sect. 3 by solving these equations  for $h(u^1, u^2)$. In the general case ($\beta \ne 0, \  \alpha+2\beta \ne 0$) we have three essentially different examples,
$$ 
h(u^1, u^2)=(u^2+ f(u^1)  )^{\frac{\beta}{\alpha+2\beta}},
$$
where $f(u^1)=a(u^1+c)^{\frac{\alpha + 2 \beta}{\beta}}+b(u^1-c)^{\frac{\alpha + 2 \beta}{\beta}}$, as well as
$$
h(u^1,u^2)=(u^1+c)^{-\frac{\alpha+\beta}{\beta}}u^2+a ({u^1-c})^{2+\alpha/\beta}({u^1+c})^{-1-{\alpha}/{\beta}}
$$
and
$$
h(u^1,u^2)=u^2+c(u^1)^2; 
$$
here   $a, b, c$ are arbitrary constants. 
In the case $\beta =0$ we have two extra examples,
$$
h(u^1,u^2)= \ln(u^2+f(u^1)), ~~~~ f''=cf,
$$
and
$$
h(u^1, u^2)=e^{cu^1}u^2+e^{2cu^1}.
$$
Finally, the case $\alpha +2 \beta=0$  gives three extra examples,
$$
h(u^1,u^2)=e^{au^2+f(u^1) },
$$
here $f(u^1)=\frac{b}{2\sqrt{b^2+1}}\ln \frac{\sqrt{b^2+1}+cu^1}{\sqrt{b^2+1}-cu^1}+\frac{1}{2}\ln (b^2+1-c^2(u^1)^2), $ as well as
$$
h(u^1, u^2)=(u^1+c)u^2+b(u^1+c)\ln \frac{u^1+c}{u^1-c}
$$
and
$$
h(u^1, u^2)=u^1u^2+\frac{c}{u^1}.
$$
These examples correct the list presented in \cite{Kup3}.

We prove in Sect. 4 that all remaining  components of the Haantjes tensor of the corresponding Hamiltonian chains vanish identically by virtue of (\ref{h}). 

In Sect. 5 we demonstrate that  the requirement of the existence of an additional conservation law of the form
$
p(u^1,u^2,u^3)_t=q(u^1,u^2,u^3,u^4)_x
$
leads to the same relations (\ref{h}).

As shown in Sect. 6,  equations (\ref{h}) imply the existence of  a generating function of conservation laws and, hence, the infinity of conservation laws. This establishes the integrability of all examples constructed in Sect. 3.  In the general case the generating function is expressed in terms of a hypergeometric function of Gauss.

\section{Derivation of the integrability conditions} 
  
There are two `trivial' cases which are to be  excluded from the  further analysis:

\noindent  (i) $\alpha+\beta=0$. In this case the Haantjes tensor vanishes identically for any Hamiltonian density $h(u^1, u^2)$, indeed,  the first two equations for $u^1, \ u^2$ form an independent subsystem, while the remaining equations  become strictly lower-triangular. 

\noindent (ii) $\triangle =(\alpha+\beta)h_2+(\alpha+2\beta)u^2h_{22}+2\beta u^1 h_{12}=0$. In this case the first equation of the chain decouples from the rest, taking the form $u^1_t=\lambda (u^1)  u^1_x$ where $\lambda (u^1)$ is a  function of $u^1$. The Haantjes tensor is also  identically zero.

\bigskip  
  
 \noindent Assuming in what follows that neither of the  expressions in (i) and (ii) vanishes (they appear as  denominators in the formulas below), calculating  the components $H^1_{jk}$ of the Haantjes tensor (one can use computer algebra to perform calculations of this type), and setting them equal to zero, we obtain the following expressions for  third order partial derivatives of the  Hamiltonian density $h(u^1, u^2)$:
\begin{eqnarray}
h_{222}&=&\frac{(2\alpha+3\beta)h_{22}^2}{(\alpha+\beta)h_2}, \nonumber \\\nonumber
\ \\
h_{122}&=&\frac{(2\alpha+3\beta)h_{22}h_{12}}{(\alpha+\beta)h_2}, \nonumber \\\nonumber
\ \\
h_{112}&=& \frac{(\alpha+2\beta)h_{12}^2+(\alpha+\beta)h_{22}h_{11}}{(\alpha+\beta)h_2}, \label{h}
\ \\
h_{111}&=&\frac{- \alpha(\alpha+2\beta)u^2 h_{12}^3 +(\alpha+\beta)((2\alpha+3\beta)h_2+3(\alpha+2\beta)u^2 h_{22})h_{11}h_{12}}{(\alpha+\beta)h_2\triangle}\nonumber\\
& &+\frac{2\beta(\alpha+3\beta)u^1 h_{12}^2h_{11}+2\alpha\beta u^1h_{22}h_{11}^2}{(\alpha+\beta)h_2\triangle};
\nonumber
\end{eqnarray}
here $\triangle=(\alpha+\beta)h_2+(\alpha+2\beta)u^2h_{22}+2\beta u^1 h_{12}$, and lower indices indicate differentiation with respect to $u^1$ and $u^2$. We have verified  the involutivity of this system.

\medskip

 \noindent {\bf Remark 1.} Since $u^1$ is the density of  the  momentum  of the Hamiltonian structure (\ref{Ham}), the addition of terms linear in $u^1$ to the Hamiltonian density $h$   effects neither the integrability of the chain, nor the vanishing of the Haantjes tensor.  Thus, the classification below is carried out up to transformations of the form 
\begin{eqnarray}
h \rightarrow ah+bu^1+c
\label{Transformation}
\end{eqnarray}  
where a, b, c are arbitrary constants.

\section{Integrable Hamiltonian densities}

We begin with the general case when both  $\beta$  and $\alpha +2 \beta $ are nonzero. The cases when either of them vanishes will be considered separately. 

\subsection {General case}

If $h_{22} \neq 0$ then, using the first two equations (\ref{h}), we obtain
$$
h_{22}= c\ h_2^{\frac{2\alpha+3\beta}{\alpha+\beta}}, \ \ \ c= const.
$$
This can be integrated twice to give $h=(u^2+f(u^1))^{\frac{\beta}{\alpha+2\beta}}+g(u^1)$.
The substitution into the third equation implies $g''=0$. Thus,  up to transformations (\ref{Transformation}), we have 
\begin{equation}
h(u^1, u^2)=(u^2+f(u^1))^{\frac{\beta}{\alpha+2\beta}}.
\label{h1}
\end{equation}
Substituting this into the remaining expression for $h_{111}$ we obtain an ODE for $f(u^1)$,
$$
f'''=\alpha f''\frac{(\alpha+\beta)f'-2\beta u^1  f''}{(\alpha+\beta)((\alpha+2\beta)f-2\beta u^1 f')},
$$
with the general solution
$$
f=a(u^1+c)^{\frac{\alpha + 2 \beta}{\beta}}+b(u^1-c)^{\frac{\alpha + 2 \beta}{\beta}};
$$
here $a, b, c$ are arbitrary constants (we thank Sasha Veselov for this observation).

If $h_{22}=0$ then the first two equations (\ref{h}) are satisfied identically, while the last two equations imply either
\begin{equation}
h(u^1,u^2)=(u^1+c)^{-\frac{\alpha+\beta}{\beta}}u^2+\tau(u^1)
\label{h3}
\end{equation}
where $\tau(u^1)$ satisfies an ODE   $\tau'''=\frac{\tau''}{\beta}\left(\frac{\alpha}{u^1-c}-\frac{\alpha+3\beta}{u^1+c}\right)$ with the general solution   
$$
\tau=a ({u^1-c})^{2+\alpha/\beta}({u^1+c})^{-1-{\alpha}/{\beta}},
$$
 or
\begin{equation}
h(u^1,u^2)=u^2+c(u^1)^2. 
\label{h2}
\end{equation}
We point out that the integrability of the Hamiltonian chain with the density (\ref{h2}) was established in \cite{Kup3}.

\medskip

\subsection{ Case $\beta=0$} 

The equations for $h$ take the form
\begin{eqnarray*}
h_{222}&=&\frac{2h_{22}^2}{h_2},\\
h_{122}&=&\frac{2h_{22}h_{12}}{h_2},\\
h_{112}&=&\frac{h_{11}h_{22}+h_{12}^2}{h_2},\\
h_{111}&=&\frac{-u^2h_{12}^3+2h_2h_{11}h_{12}+3u^2h_{22}h_{12}h_{11}}{h_2(h_2+u^2h_{22})}.
\end{eqnarray*}
\noindent If $h_{22}\ne 0$ then,  up to transformations  (\ref{Transformation}), the first three equations imply
\begin{equation}
h(u^1,u^2)= \ln(u^2+f(u^1)),
\label{h4}
\end{equation}
and the substitution of this ansatz into the fourth equation gives $f'''f-f'f''=0$, which  integrates to  $f''=cf, \ c= const.$ If $h_{22}=0$ then the first two equations (\ref{h}) are satisfied identically, while the other two imply either
\begin{equation}
h(u^1, u^2)=e^{cu^1}u^2+e^{2cu^1},
\label{h5}
\end{equation}
 or
\begin{equation}
h(u^1, u^2)=u^2+c(u^1)^2,
\label{h22}
\end{equation}
$c=const$, the latter case coinciding with (\ref{h2}).

\medskip

\subsection{Case $\alpha+2\beta=0$ } 

The equations for $h$ take the form
\begin{eqnarray*}
h_{222}&=&\frac{h_{22}^2}{h_2},\\
h_{122}&=&\frac{h_{22}h_{12}}{h_2},\\
h_{112}&=&\frac{h_{11}h_{22}}{h_2},\\
h_{111}&=&\frac{h_2h_{11}h_{12}+2u^1h_{12}^2h_{11}-4u^1h_{22}h_{11}^2}{h_2(h_2-2u^1h_{12})}.
\end{eqnarray*}
If $h_{22}\ne 0$ then, up to transformations  (\ref{Transformation}), the first three equations imply
\begin{equation}
h(u^1,u^2)=e^{au^2+f(u^1) }.
\label{h6}
\end{equation}
The substitution of  this ansatz into the fourth equation gives an ODE for $f$,
\begin{eqnarray*}
f'''+2f'f''=2u^1(f'f'''-2(f'')^2),
\end{eqnarray*} 
which can be solved as follows. Setting $f'=1/w$ we obtain 
$w''(w-2u^1)=2w'(w'-1)$. Introducing $w=v+2u^1$ we arrive at an autonomous ODE 
$v''v=2(v'+2)(v'+1)$ which integrates twice leading  to the general solution
$$
f=\frac{b}{2\sqrt{b^2+1}}\ln \frac{\sqrt{b^2+1}+cu^1}{\sqrt{b^2+1}-cu^1}+\frac{1}{2}\ln (b^2+1-c^2(u^1)^2), ~~~ c, b =const,
$$
or its linear degeneration $f=cu^1$. Setting $b=\sinh a$ we can rewrite the above expression in the form
$$
f=\frac{1}{2\cosh a}\left( e^a\ln (\cosh a+cu^1)+ e^{-a}\ln (\cosh a-cu^1)\right).
$$
The case $h_{22}=0$ leads to either
\begin{equation}
h(u^1, u^2)=(u^1+c)u^2+b(u^1+c)\ln \frac{u^1+c}{u^1-c},
\label{h7}
\end{equation}
where $c,  \ b$ are arbitrary constants, or its degeneration,
\begin{equation}
h(u^1, u^2)=u^1u^2+\frac{c}{u^1}.
\label{h8}
\end{equation}
Further degeneration $h_{22}=h_{12}=0$ results in
\begin{equation}
h(u^1, u^2)=u^2+c(u^1)^2.
\label{h222}
\end{equation}

\section{The vanishing of the Haantjes tensor}

In this section we demonstrate that the relations (\ref{h}), which were obtained from the requirement of the vanishing of the first few components $H^1_{jk}$ of the Haantjes tensor, are already sufficiently restrictive and imply the vanishing of all other components. The proof utilizes an important property of Hamiltonian chains (\ref{Ham}), namely, the existence of finite-component reductions for {\it any} (not necessarily integrable) Hamiltonian density $h$ \cite{Maks1}. Let us parametrize $u^i$ in terms of finitely many `moments' $v^a, \ a=1,..., n$, as follows:
$$
u^k=\frac{1}{2+\frac{\alpha}{\beta}(k-1)}\sum_1^n(v^a)^{2+\frac{\alpha}{\beta}(k-1)},
$$
$k=1, 2, 3, ....$ (we consider the generic case when all expressions in the denominators are non-vanishing, see \cite{Maks1} for a discussion of the exceptional cases). Thus, 
\begin{equation}
u^1=\frac{1}{2}\sum_1^n(v^a)^2, ~~~ u^2=\frac{1}{2+\frac{\alpha}{\beta}}\sum_1^n(v^a)^{2+\frac{\alpha}{\beta}},
\label{u12}
\end{equation}
etc. One can verify that under this substitution the infinite chain (\ref{Ham}),
$$
u^k_t=(\alpha(k-1)+\beta)u^k(h_1)_x+(\alpha(k-1)+\beta)u^{k+1}(h_2)_x+\beta(u^kh_1)_x+
(\alpha+ \beta)(u^{k+1}h_2)_x,
$$
 reduces to an $n$-component `symmetric' conservative system for $v^a$, 
\begin{equation}
v^a_t=\beta\left(v^ah_1+(v^a)^{1+\frac{\alpha}{\beta}}h_2\right)_x,
\label{red}
\end{equation}
$a=1, ..., n$, see \cite{Maks2}. Here $h(u^1, u^2)$ is an {\it arbitrary} Hamiltonian density, not necessarily satisfying the integrability conditions (\ref{h}), and $u^1, u^2$ are given by (\ref{u12}). Notice that the system (\ref{red}) is manifestly Hamiltonian: $v^a_t=\beta(\partial h/\partial {v^a})_x.$ Similar formulae can be obtained if the density $h$ depends on more than two $u$'s. We have the following

\begin{theorem}
For a Hamiltonian density $h(u^1, u^2)$ satisfying the integrability conditions (\ref{h}),  the Haantjes tensor of any finite-component reduction (\ref{red}) is identically zero.
\end{theorem}

The proof follows from the  explicit formulas for the Nijenhuis tensor,
\begin{eqnarray*}
&&N^1_{23}=\beta v^1v^2v^3((v^2)^{\frac{\alpha} {\beta}}-(v^3)^{\frac{\alpha}{\beta}})\left[ 2\beta^2 u^1(h_{11}h_{12}^2-h_{11} ^2h_{22}) +\beta(2+\frac{\alpha}{\beta})(\alpha u^2-2\beta (v^1)^{\frac{\alpha}{\beta}} u^1)(h_{11}h_{22}h_{12}-\right. \\
&& \left. h_{12}^3) +(\alpha+\beta)(2(1+\frac{\alpha}{\beta})u^3-\beta (2+\frac{\alpha}{\beta})^2(v^1)^{\frac{\alpha}{\beta}} u^2)(h_{11}h_{22}^2-h_{12}^2h_{22}) + (\alpha+\beta)h_2   [(\alpha+2\beta) \right. \\
&& \left.(h_{11}h_{12}+(v^1v^2v^3)^{\frac{\alpha}{\beta}}h_{22}^2)   +((v^1v^2)^{\frac{\alpha}{\beta}}+(v^1v^3)^{\frac{\alpha}{\beta}}+(v^2v^3)^{\frac{\alpha}{\beta}})h_{12}h_{22} +\right.\\
&&\left. ((\alpha+\beta)((v^1)^{\frac{\alpha}{\beta}}+(v^2)^{\frac{\alpha}{\beta}}+(v^3)^{\frac{\alpha}{\beta}})-(2\alpha + \beta)((v^1)^{\frac{\alpha}{\beta}})h_{11}h_{22} + (\beta((v^1)^{\frac{\alpha}{\beta}}+(v^2)^{\frac{\alpha}{\beta}}+(v^3)^{\frac{\alpha}{\beta}})\right.\\
&&\left.+(2\alpha + \beta) (v^1)^{\frac{\alpha}{\beta}})h_{12}^2)]\right ]
\end{eqnarray*}
and
\begin{eqnarray*}
&&N^1_{12}=\beta^2(v^1)^2v^2((v^1)^{\frac{\alpha}{\beta}}-(v^2)^{\frac{\alpha}{\beta}})\left[(2+\frac{\alpha}{\beta})(\alpha u^2-2\beta (v^1)^{\frac{\alpha}{\beta}} u^1)(h_{11}h_{22}h_{12}-h_{12}^3)+\right.\\
&&\left.2\beta^2 u^1(h_{11}h_{12}^2-h_{11} ^2h_{22})+((2+\frac{\alpha}{\beta})(\alpha-2\beta)(v^1)^{\frac{\alpha}{\beta}}u^2-2(\alpha+\beta)(1+\frac{\alpha}{\beta})u^3)(h_{12}^2h_{22}-h_{22}^2h_{11})\right]-\\
&&\beta(\alpha+\beta)h_2v^2\left[(v^1v^2)^{\frac{\alpha}{\beta}}(2(\alpha+\beta)(1+\frac{\alpha}{\beta})u^3+(\alpha+2\beta)((v^2)^{\frac{\alpha}{\beta}}-(v^1)^{\frac{\alpha}{\beta}})(v^1)^{2+\frac{\alpha}{\beta}})h_{22}^2+\right.\\
&&\left.(-\alpha(v^1)^{\frac{\alpha}{\beta}}(-(2+\frac{\alpha}{\beta})u^2-2(v^2)^{\frac{\alpha}{\beta}}u^1+2((v^1)^{\frac{\alpha}{\beta}}-(v^2)^{\frac{\alpha}{\beta}})(v^1)^2) +\beta((2+\frac{\alpha}{\beta})\right.\\
&&\left.((v^1)^{\frac{\alpha}{\beta}}+(v^2)^{\frac{\alpha}{\beta}})u^2+2(v^1v^2)^{\frac{\alpha}{\beta}}u^1+2(1+\frac{\alpha}{\beta})u^3+(v^1)^{2+\frac{\alpha}{\beta}}(2(v^2)^{\frac{\alpha}{\beta}}-3(v^1)^{\frac{\alpha}{\beta}})+(v^2)^{\frac{\alpha}{\beta}}(v^1)^2))h_{12}^2\right.\\ 
&&\left.+2\beta u^1 h_{11}^2+(2((\alpha+\beta)(v^1)^{\frac{\alpha}{\beta}}+\beta (v^2)^{\frac{\alpha}{\beta}})u^1+\beta(2+\frac{\alpha}{\beta})(((v^2)^{\frac{\alpha}{\beta}}-(v^1)^{\frac{\alpha}{\beta}})(v^1)^2+u^2))h_{11}h_{12} \right.\\
&&\left.+(2(1+\frac{\alpha}{\beta})((\alpha+\beta)(v^1)^{\frac{\alpha}{\beta}}+\beta(v^2)^{\frac{\alpha}{\beta}})u^3+(2+\frac{\alpha}{\beta})(v^1)^{\frac{\alpha}{\beta}}(2(v^2)^{\frac{\alpha}{\beta}}(\alpha+\beta)u^3+\beta(2(v^2)^{\frac{2\alpha}{\beta}}-(v^1)^{\frac{2\alpha}{\beta}}-\right.\\ 
&&\left. (v^1v^2)^{\frac{\alpha}{\beta}})(v^1)^2))h_{12}h_{22}\right]-(\alpha+\beta)^2h_2v^2\left[((\alpha-\beta)(v^1)^{\frac{\alpha}{\beta}}+\beta(v^2)^{\frac{\alpha}{\beta}})h_{11}+\right. \\ 
&&\left. ((\alpha+\beta)(v^2)^{\frac{\alpha}{\beta}}-\beta (v^1)^{\frac{\alpha}{\beta}})(v^1v^2)^{\frac{\alpha}{\beta}}h_{22}+(2\alpha(v^1v^2)^{\frac{\alpha}{\beta}}+\beta((v^2)^{\frac{2\alpha}{\beta}}-(v^1)^{\frac{2\alpha}{\beta}}))h_{12}\right],
\end{eqnarray*}
which were obtained using the integrability conditions (\ref{h}). Notice that,  since (\ref{red}) is invariant under  permutations of $v$'s, it is sufficient to specify  $N^1_{23}$ and $N^1_{12}$ only. We emphasize that these expressions do not explicitly depend on the size $n$ of the reduction (\ref{red}): this dependence is hidden in the variables $u^1, u^2, u^3$. The vanishing of the Haantjes tensor (\ref{H}) is now a straightforward algebraic calculation.

Thus, for the Hamiltonian densities $h(u^1, u^2)$ which satisfy the integrability conditions (\ref{h}), the corresponding hydrodynamic chains possess   diagonalizable $n$-component reductions (\ref{red}) for any value of $n$. According to the results of \cite{FerMar},  this implies that the full Haantjes tensor of the Hamiltonian chain vanishes identically.

\section{Conservation laws}

For any Hamiltonian density $h(u^1, u^2)$ the system (\ref{Ham}) necessarily possesses two conservation laws, namely,
\begin{eqnarray*}
u^1_t&=& (2\beta u^1h_1+(\alpha+2\beta)u^2h_2-\beta h)_x
\end{eqnarray*}
and
\begin{eqnarray*}
h_t &=&((\alpha+\beta)u^3h_2^2+(\alpha+2\beta)u^2h_1h_2+\beta u^1h_1^2)_x,
\end{eqnarray*}
which correspond to the conservation of the momentum and the Hamiltonian, respectively. Let us require that there exists an `extra'  conservation law of the form
\begin{eqnarray}
p(u^1,u^2,u^3)_t=q(u^1,u^2,u^3,u^4)_x.
\label{conv}
\end{eqnarray}

\begin{theorem}
The integrability conditions (\ref{h}) are necessary and sufficient for the existence of an additional conservation law  of the form (\ref{conv}).
\end{theorem}

The proof is computational: substituting in for $u^1_t, u^2_t, u^3_t, u^4_t$  into the left hand side of (\ref{conv}), we collect coefficients at $u^1_x, ..., u^4_x$ and equate them to zero.  This results in a system of first order partial differential equations for the flux $q$, 
\begin{eqnarray*}
q_1&=& (3\alpha+2\beta)u^4p_3h_{12} +(\alpha+2\beta)u^2(p_1h_{12}+p_2h_{11})+2(\alpha+\beta)u^3(p_3h_{11}+p_2h_{12})+\beta h_1p_1+\\
& & 2\beta u^1 p_1h_{11}, \\ 
q_2&=&(3\alpha+2\beta)u^4h_{22}p_3+(\alpha+\beta)(2u^3(h_{12}p_3+h_{22}p_2)+h_2p_1)+(\alpha u^2p_2+2\beta (u^1p_1+u^2p_2))h_{12}+\\
& & u^2(\alpha+2\beta)p_1h_{22}+\beta h_1p_2,\\
q_3&=&\beta h_1 p_3+(\alpha+\beta)h_2p_2,\\ 
q_4&=&(\alpha+\beta)h_2p_3.
\end{eqnarray*}
Calculating the consistency conditions for the flux, $q_{ij}=q_{ji}$, we obtain all   second order partial derivatives of the density $p$,  
\begin{eqnarray*}
p_{11} &=& \frac{2(2\alpha+\beta)((\alpha+\beta)u^3h_{12}^2+\beta u^1h_{11}^2+(\alpha+2\beta)u^2h_{12}h_{11})p_3 +(\alpha+\beta)^2 p_2 h_2h_{11}}{(\alpha+\beta)^2 h_2^2}, \\
p_{12} &=&\frac{(\alpha+\beta)^2h_2h_{12}p_2+(2\alpha+\beta)((\alpha+\beta)(2u^3h_{12}+h_2h_{22})+2\beta h_{11}h_{12}+(\alpha+2\beta)(h_{12}^2+h_{11}h_{22}))p_3}{(\alpha +\beta)^2 h_2^2},\\
p_{22}&=&\frac{(2\alpha+\beta)((\alpha+\beta)(2u^3p_3h_{22}^2+2h_{12}h_2p_3)+2((\alpha+2\beta)u^2h_{22}+\beta u^1h_{12})p_3h_{12})+(\alpha+\beta)^2h_{22}h_2p_2}{(\alpha+\beta)^2h_2}, \\
p_{13}&=&\frac{(2\alpha+\beta)h_{12}p_{3}}{(\alpha+\beta)h_2},\\
p_{23}&=&\frac{(2\alpha+\beta)h_{22}p_{3}}{(\alpha+\beta)h_2},\\
p_{33}&=&0. 
\end{eqnarray*}
These equations imply, in particular, that $p$ is linear in $u^3$, and $q$ is linear in $u^4$. Upon imposing the consistency condition $(p_{ij})_k=(p_{ik})_j$ we get a system of third order partial differential equations for the density $h(u^1,u^2)$, which is identical to (\ref{h}). This finishes the proof.

We will demonstrate in Sect. 6 that  equations (\ref{h}) imply the existence of a generating function for  conservation laws. Thus,  we can claim that the existence of  one additional conservation law is already very restrictive and implies the existence of an  infinity of conservation laws, thus manifesting the integrability.

{\bf Remark.} Taking $p(u^1, u^2, u^3)$ defined by the above equations as a Hamiltonian density, one obtains a  system (\ref{Ham}) which commutes with the Hamiltonian system defined by the density $h(u^1, u^2)$, and  also has the identically vanishing Haantjes tensor. We emphasize that the densities $p(u^1, u^2, u^3)$ arising in this way are necessarily linear in $u^3$. One can formulate a natural question: describe all Hamiltonian densities of the form $p(u^1, u^2, u^3)$ such that the associated  chain (\ref{Ham}) has the vanishing Haantjes tensor. A detailed analysis of this problem leads to the two possibilities:

\noindent (i) the density $p$ is linear in $u^3$. In this case one can show the existence of a lower order commuting flow with the  Hamiltonian density $h(u^1, u^2)$, which brings us back to the situation discussed above.

\noindent (ii) the density $p$ is such that the quantity $2\beta u^1p_1+(\alpha+2\beta)u^2p_2+2(\alpha + \beta)u^3p_3-\beta p$ is a function of the first coordinate $u^1$ only. In this case the first equation of the chain,
$$
u^1_t=(2\beta u^1p_1+(\alpha+2\beta)u^2p_2+2(\alpha + \beta)u^3p_3-\beta p)_x,
$$
decouples from the rest. The general form of all such densities  is
$$
p=\sqrt {u^1}F\left({u^2}{(u^1)^{-\frac{\alpha+2\beta}{2\beta}}}, \ {u^3}{(u^1)^{-\frac{\alpha+\beta}{\beta}}}\right)+f(u^1)
$$
where $F$ and $f$ are arbitrary functions. 
This result shows that, essentially,  there exist no nontrivial `genuine'  integrable densities of the form $p(u^1, u^2, u^3)$.

\section{Generating functions for conservation laws}

The structure of   reductions (\ref{red}) suggests that one should seek a generating function for conservation laws in the form 
\begin{equation}
\lambda = \lambda (p, u^1, u^2, u^3,...)
\label{lambda}
\end{equation}
so that the following Gibbons-type relation holds:
\begin{equation}
\lambda_t-\beta \left(h_1+(1+{\alpha}/{\beta})p^{\frac{\alpha}{\beta}}h_2\right ) \lambda_x=\lambda_p\left(p_t-\beta(ph_1+p^{1+\frac{\alpha}{\beta}}h_2)_x\right);
\label{genfun}
\end{equation}
this relation is required to be satisfied identically modulo  (\ref{Ham}). We will demonstrate the existence of a generating function of this form  for any Hamiltonian density satisfying the integrability conditions (\ref{h}). Suppose the relation (\ref{genfun}) is already established.
Then, setting $\lambda=const$, one has $\lambda_t=\lambda_x=0$, so that the relation (\ref{genfun}) takes the form
$$
p_t-\beta\left(ph_1+p^{1+\frac{\alpha}{\beta}}h_2\right)_x=0.
$$
This provides an infinite sequence of conserved densities after one expands $p$ as a series in $\lambda$ by virtue of (\ref{lambda}). The method of generating functions is standard, and can be traced back to \cite{Benney, Gibb81, Kup1, Kup2}, see also \cite{Maks1} for recent developments. A detailed analysis of the relation (\ref{genfun}) reveals that  the dependence of $\lambda$ on $u^2, u^3, u^4, $ etc, is fixed uniquely,
\begin{equation}
\lambda=\sum_{k=2}^{\infty} q^{1-2\frac{\beta}{\alpha}-k}\ u^k+s(u^1, q), ~~~ q=p^{\frac{\alpha}{\beta}},
\label{sum}
\end{equation}
while the function $s(u^1, q)$, which  specifies the dependence of $\lambda$ on $u^1$, satisfies the equations
\begin{equation}
s_1=q^{-1-2\frac{\beta}{\alpha}}F, ~~~~
s_q=\frac{1}{\alpha}q^{-1-2\frac{\beta}{\alpha}}G.
\label{s}
\end{equation}
Here $F$ and $G$ are the following rational expressions in $q$ depending on the Hamiltonian density $h$:
$$
\begin{array}{c}
F=\frac{(h_{11}+qh_{12})((\alpha + \beta)qh_2-(\alpha+2\beta)u^2h_{12})+(\alpha+2\beta)u^2h_{11}(h_{12}+qh_{22})}
{(\alpha+\beta)h_2[q^2h_{22}+2 q h_{12}  +h_{11}]+(2\beta q u^1 -(\alpha+2\beta)u^2)[h_{12}^2-h_{11}h_{22}]}, \\
\ \\
G=\frac{(\alpha + \beta)^2q h_2 +2 \beta (\alpha+2\beta)u^1 u^2 (h_{12}^2-h_{11}h_{22})+(\alpha+\beta)h_2 [(\alpha+2\beta)q  u_2 h_{22}-2\beta u^1 h_{11}]}{(\alpha+\beta)h_2[q^2h_{22}+2q h_{12}  +h_{11}]+(2\beta q u^1 -(\alpha+2\beta)u^2)[h_{12}^2-h_{11}h_{22}]}.
\end{array}
$$
With the ansatz (\ref{sum}), all terms in (\ref{genfun}) containing $p_t,  p_x,  u^3_x, u^4_x, ...$,  cancel identically, while the requirement of cancellation of the coefficients at $u^1_x$ and $u^2_x$ results in (\ref{s}).
The functions $F$ and $G$ satisfy the relations
\begin{equation}
G_1=\alpha F_q-(\alpha+2\beta)q^{-1} F \label{FG}, ~~~ F_2=G_2=0,
\end{equation}
which are the consistency conditions of the equations (\ref{s}). These relations are  satisfied identically modulo the integrability conditions (\ref{h}).  Conversely,  the relations (\ref{FG})   imply the integrability conditions (\ref{h}).
 For each of the cases arising in the classification in Sect. 3, the equations (\ref{s}) for $s(u^1, q)$ can be solved explicitly. Thus, the generating function $\lambda$ can be reconstructed in closed form. This is mainly due to a simple dependence of the derivative $s_q$ on $q$: integrating it with respect to $q$ we obtain a closed form expression which, in most of the cases, automatically solves the equation for $s_1$. We consider the canonical forms
(\ref{h1})--(\ref{h222}) case by case below.

\subsection{Generating functions in the general case: densities (\ref{h1})--(\ref{h2})}

For the Hamiltonian density (\ref{h1}) the function $s(u^1, q)$ is given by 
$$
s(u^1, q)=\frac{\beta^2q^{1-2\beta/\alpha}}{4abc^2(\alpha^2-4\beta^2)}\Big[
a(c-u^1)^{2-\frac{\alpha}{\beta}}F\left(\frac{\beta q(u^1-c)^{-\alpha/\beta}}{2bc(\alpha+2\beta)}\right) 
$$
$$
+b((c+u^1)^{2-\frac{\alpha}{\beta}}F\left(-\frac{\beta q(u^1+c)^{-\alpha/\beta}}{2ac(\alpha+2\beta)}\right)\Big]+q^{-2\beta/\alpha}u^1;
$$
here $F(t)={}_2F_1(1-2\beta/\alpha, \ 1, \ 2-2\beta/\alpha, \ t)$ is the hypergeometric function of Gauss. In the case (\ref{h3}) one has
$$
s(u^1, q)=\frac{\beta^2q^{1-2\beta/\alpha}(u^1-c)^{2-\alpha/\beta}}{4ac^2(\alpha^2-4\beta^2)}
F\left(\frac{\beta q(u^1-c)^{-\alpha/\beta}}{2ac(\alpha+2\beta)}\right) 
+q^{-2\beta/\alpha}u^1;
$$
here $F$ is the same hypergeometric function as above. Finally, the case of (\ref{h2}) leads to
\begin{equation}
s(u^1, q)=\left(u^1+\frac{\alpha+\beta}{2c(\alpha-2\beta)}q\right)q^{-2\beta/\alpha}.
\label{ss}
\end{equation}

\subsection{Generating functions in the case $\beta=0$: densities (\ref{h4})--(\ref{h22})}

Here the relation (\ref{genfun}) takes the form 
$$
\lambda_t-e^ph_2 \lambda_x=\lambda_p\left(
p_t-(h_1+e^ph_2)_x\right)
$$
where 
$$
\lambda=\sum_{k=2}^{\infty} q^{1-k}\ u^k+s(u^1, q), ~~~ q=e^p.
$$
The function $s(u^1, q)$  satisfies the equations
$$
\begin{array}{c}
s_1=\frac{h_2(h_{11}+qh_{12})+u^2(h_{11}h_{22}-h_{12}^2)}
{u^2(h_{11}h_{22}-h_{12}^2)+h_2(h_{11}+2 q h_{12}+q^2h_{22})}, \\
\ \\
s_q= \frac{h_2(u^2h_{22}+h_{2})}{u^2(h_{11}h_{22}-h_{12}^2)+h_2(h_{11}+2 q h_{12}+q^2h_{22})},
\end{array}
$$
which are consistent and define $s(u^1, q)$ explicitly: for the Hamiltonian density (\ref{h4}) one has
$$
s(u^1, q)=\frac{1}{\sqrt {c}} {\rm Arctanh} \left(\frac{q+f'}{\sqrt {c} f}\right);
$$
recall that $f''=cf$. The  density (\ref{h5}) leads to
$$
s(u^1, q)=\frac{1}{2}u^1+\frac{1}{2c}\log (2ce^{cu^1}+q).
$$
The density (\ref{h22}) gives
$$
s(u^1, q)=u^1+q/2c;
$$
notice that this expression can be obtained from (\ref{ss}) by setting $\beta=0$.

\subsection{Generating functions in the case $\alpha+2\beta=0$: densities (\ref{h6})--(\ref{h222})}

Here the relation (\ref{genfun}) takes the form (set $\beta=1, \alpha=-2$):
$$
\lambda_t-\left(h_1-p^{-2}h_2\right ) \lambda_x=\lambda_p\left(
p_t-(h_1p+p^{-1}h_2)_x\right)
$$
where 
$$
\lambda=\sum_{k=2}^{\infty} q^{2-k}\ u^k+s(u^1, q), ~~~ q=p^{-2}.
$$
The function $s(u^1, q)$  satisfies the equations
$$
\begin{array}{c}
s_1=-qh_2\frac{h_{11}+qh_{12}}
{(h_{11}+qh_{12})(2u^1h_{12}-h_2)-(h_{12}+qh_{22})(2u^1h_{11}+qh_2)}, \\
\ \\
s_q=-\frac{h_2}{2} \frac{2u^1h_{11}+qh_2}{(h_{11}+qh_{12})(2u^1h_{12}-h_2)-(h_{12}+qh_{22})(2u^1h_{11}+qh_2)},
\end{array}
$$
which are consistent and define $s(u^1, q)$ explicitly: for the Hamiltonian density (\ref{h6}) one has
$$
s(u^1, q)=-\frac{b{\rm Arctanh}\left(\frac{cu^1}{\sqrt {1+b^2}}\right)}{2\sqrt{1+b^2}}+
\frac{b{\rm Arctanh}\left(\frac{2q(1+b^2-c^2(u^1)^2)^2-cb(b-cu^1)^2-cb+2c^2u^1}{c\sqrt{1+b^2}(1+(b-cu^1)^2)}\right)}{4\sqrt{1+b^2}}
$$
$$
+\frac{1}{8}\log \big[-4cbq(1+b^2)+4q^2(1+b^2)^2-4c^3bq(u^1)^2+4c^4q^2(u^1)^4-c^2+8c^2qu^1(1+b^2)(1-qu^1)
\big].
$$
The  density (\ref{h7}) leads to
$$
s(u^1, q)=\frac{1}{4c}q(u^1+c)^2+\frac{b}{2}\big[1-\log(q(u^1-c)^2+2cb)\big].
$$
In the case of (\ref{h8}) one has
$$
s(u^1, q)=qu^1+\frac{1}{8c}q^2(u^1)^4.
$$
Finally, the case of (\ref{h222}) gives
$$
s(u^1, q)=qu^1+\frac{1}{8c}q^2;
$$
this follows from (\ref{ss}) when $\beta=1, \alpha=-2$.

\medskip

We point out that the case $\alpha=0$ requires a special treatment: it corresponds to the chains which possess linearly degenerate hydrodynamic reductions. The  approach of generating functions does not apply to this class, see e.g.  \cite{Shabat1} for a discussion of a particular example of this type.

\section{Concluding remarks}

There exists a whole variety of approaches to the classification of integrabile  Hamiltonian hydrodynamic chains, based on seemingly different requirements, namely:

\noindent --- the vanishing of the Haantjes tensor;

\noindent --- the existence of infinitely many $n$-component hydrodynamic reductions for any $n$;

\noindent --- the existence of one `extra' conservation law.

\noindent --- the existence of a generating function for conservation laws;

All examples known to us support the evidence that these approaches are essentially equivalent, leading to the same integrability conditions and classification results. Among others, the approach based on the Haantjes tensor seems to be the most universal, leading to the required integrability conditions in a  straightforward way. We emphasize that  components of the Haantjes tensor can be calculated using computer algebra. Moreover,  the vanishing of the first few components  $H^1_{jk}$ is already sufficiently restrictive and implies the identical vanishing of the Haantjes tensor. 

It should be pointed out that, at present,  there is no hope to {\it prove} the equivalence of the approaches listed above in the full generality: one first needs a classification of infinite-dimensional Poisson brackets of hydrodynamic type, which is an interesting and non-trivial problem in its own \cite{Maks3}. Some
approaches in this directions were outlined recently in \cite{Maks1}.

\section*{Acknowledgements}
The work of EVF has been partially supported by the European Union through the FP6 
Marie Curie RTN project {\em ENIGMA} (Contract number MRTN-CT-2004-5652). We thank the London Mathematical Society for their support of MVP to Loughborough, making this collaboration possible.

\end{document}